\documentclass[11pt]{article}
\usepackage{mathrsfs}
\usepackage{amsmath}

\global\arraycolsep=1pt \oddsidemargin.20in \evensidemargin.5in
\usepackage[Symbol]{upgreek}

\usepackage{bbm}
\usepackage{dsfont}
\usepackage{amssymb}
\usepackage{textcomp}

\DeclareMathAlphabet{\mathpzc}{OT1}{pzc}{m}{it}
\def\s{\,\mathpzc{s}\,}
\def\s{s}

\usepackage[colorlinks=true,backref=true,
linkcolor=black,anchorcolor=black,citecolor=black,
filecolor=black,menucolor=black,pagecolor=black,urlcolor=black]{hyperref}

\def\z{\zeta}

\def\gauge{\delta^{\scriptscriptstyle  \,gauge}}

\def\be{\begin{equation}}
\def\ee{\end{equation}}
\def\bea{\begin{eqnarray}}
\def\eea{\end{eqnarray}}
\def\bdis{\begin{displaymath}}
\def\edis{\end{displaymath}}
\def\corr{$\clubsuit$}
\def\nn{\nonumber}

\begin{document}
\allowdisplaybreaks[1]
\renewcommand{\thefootnote}{\fnsymbol{footnote}}
\def\corr{$\spadesuit $}
\def\trefle{ $\clubsuit$}
\renewcommand{\thefootnote}{\arabic{footnote}}
\setcounter{footnote}{0}
 \def\stop{$\blacksquare$}
\begin{titlepage}
\null
\begin{flushright}
CERN-PH-TH/2010 \\
{CBPF-NF-002/10}
\end{flushright}
\begin{center}
{{\Large \bf
One-dimensional structures behind \\ twisted and   untwisted superYang-Mills theory
}}
\lineskip.75em \vskip 3em \normalsize {\large Laurent
Baulieu$^{ \dagger\ddagger}$\footnote{email address:
baulieu@lpthe.jussieu.fr}
and
 Francesco Toppan$^{*}$\footnote{email address:
 toppan@cbpf.br}
\\
\vskip 1em
  $^{\dagger}${\it Theoretical Division CERN }\footnote{ CH-1211
  Gen\`eve, 23,
  Switzerland }
\\
$^{\ddagger}${\it LPTHE
Universit\'e Pierre et Marie Curie }\footnote{ 4
place Jussieu,
F-75252 Paris
Cedex 05, France}
\\
$^{* }${\it CBPF, Rio de Janeiro }\footnote{ {Rua Dr. Xavier
Sigaud 150, cep 22290-180, Rio de Janeiro, RJ, Brazil}}
 }

\vskip 1 em
\end{center}
\vskip 1 em
\begin{abstract}
~\\
We give a one-dimensional interpretation of the four-dimensional
twisted $N=1$ superYang-Mills theory on a K\"ahler manifold by
performing an appropriate dimensional reduction. We prove the existence of a
$6$-generator superalgebra, which
does not possess any invariant Lagrangian
but contains two different subalgebras that determine
the twisted and untwisted formulations of the $N=1$ superYang-Mills theory.
 \end{abstract}
\end{titlepage}

\section{Introduction.}

In this paper we investigate the relation between untwisted (Poincar\'e) and twisted
supersymmetry from a  one-dimensional point of view. We use as an example
both the vector and the scalar supermultiplets of the $N=1$, $D=4$ twisted
superYang-Mills theory. They are
defined in terms of the $SU(2)$-invariant decompositions, that
can be done on a K\"ahler manifold, of the spinors.
The one-dimensional results are basically obtained by a dimensional reduction
and an appropriate gauge-fixing,
and then reinterpreted from a one-dimensional superalgebra viewpoint.
This work is partially motivated by the intriguing question of understanding
the process of oxidizing to higher dimensions the rich supersymmetry
structure that can
be systematically obtained in one dimension. It also sheds new light to
the meaning of the twist   for supersymmetric theories.

The sets of one-dimensional
supersymmetry generators $Q_I$ can be presented
in a way that their non-vanishing anticommutators
are only present on the diagonal,
\bea\label{gensusy}
\{ Q_I,Q_J\} &=& \eta_{IJ}H,\nonumber\\
\relax [H, Q_I]&=& 0.
\eea
The classification of the supersymmetric representations for $\eta_{II}= \pm 1$
has been the subject, following \cite{pt}, of many publications \cite{{krt},{dfghil},{dfghil2},{kt},{kt2},{gkt}}. It is derived from
the classification \cite{abs} (see also \cite{{oku},{crt}}) of Clifford algebras of (non-degenerate) $(p,q)$ signature.
Here we point out that the
dimensionally reduced
$N=1$, $D=4$ twisted theory implies that $\eta_{II}$ can be possibly zero,
so that more
effort should be done for understanding systems with a general
signature ${\cal N} \equiv [n_+, n_-,n_0]$, with $n_0\neq 0$
($n_\pm, n_0$ denotes the number of terms on the diagonal that are respectively $\pm 1$ and $0$
and, for $n_-=n_0=0$, we recover the standard supersymmetry algebra of   supersymmetric quantum mechanics \cite{wit}).
With this generalization we hope that one can describe all possible twisted and untwisted supersymmetric theories.
The classification of non-trivial representations of Clifford algebras with extra anticommuting
Grassmann-type parameters  and their associated superalgebras is not contained in \cite{abs} and, respectively, in \cite{pt}.
In the absence of a general classification we prove, however, the existence of a specific non-trivial realization, as a result of
a dimensional reduction of the twisted $N=1$, $D=4$ supersymmetry.\par
The gauge symmetry is a non-trivial feature in higher dimensions,
where one usually builds equivariant supersymmetry algebra and the
balance between commuting and anticommuting
fields  is achieved modulo
gauge symmetries. One can on the other hand introduce new fields,
called shadow fields, such that their transformation laws compensate
for the appearance of the
gauge
transformations in the closure of supersymmetry relations \cite{BG}. It is
however more illuminating, in view of understanding the correspondence
with the simplest
possible one-dimensional
supersymmetric structures, to perform an appropriate gauge-fixing of the
Yang-Mills symmetry, which allows one to reach an exact balance between
bosons and fermions
without introducing shadows. This gauge-fixing automatically
implies a formulation in lower-dimensions,
where one learns new features about the twist operation.\par
Our aim is thus to understand which phenomena occur when the twisted
and untwisted
four dimensional Yang-Mills supersymmetry are projected to lower dimensions.
We will briefly recall the main features of the twisted $N=1$ Poincar\'e
supersymmetry, which
implies using the holomorphic and antiholomorphic decompositions of the
spinors on K\"ahler
manifolds. Then we will start the dimensional reduction by
setting one of the gauge field components equal to zero, which implies a
dimensional reduction to $3$ dimensions, if one preserves a $3$-generator
superalgebra
and a dimensional reduction to $1$ dimension, if a $4$-generator
superalgebra is preserved.
In one dimension we obtain representations of the
${\cal N}_{superPoinc.} \equiv [4,0,0]$ and
${\cal N}_{topol.} \equiv [1,1,2]$ superalgebras.  They are  inequivalent,
from a standard algebraic point of view. However,
in the language of  the path integral,
both representations  are related by a complexification, followed by a linear mapping and a reality
condition. If moreover one considers their $3$-generator subalgebras
$[3, 0,0]\subset  [4, 0,0]$
and $[1, 1,1]\subset [1, 1,2]  $, which are big enough to determine the Lagrangians
for both theories,
they end up to be different truncations of a $6$-generator algebra $[3, 3,0]$.
The latter algebra itself is too big to admit an invariant Lagrangian.
Altogether,
the structure that we found in this analysis is rich enough to deserve an exposition.
Our results have generalizations for extended supersymmetry. Moreover, the oxidation of
the  one-dimensional algebra to $2n$ dimensions by imposing a $SU(n)$ global invariance
seems a promising method \cite{bln}.

\def\z{\bar z}
\def\bm{{\bar m}}
\def\bn{{\bar n}}
\def\bp{{\bar p}}
\def\bq{{\bar q}}

\section{The $N=1$, $D=4$ vector and scalar multiplets and
their twisted symmetries.}

The $N=1$ vector multiplet of the $N=1$, $D=4$ superPoincar\'e theory
contains one gauge field, one Majorana spinor and one auxiliary scalar boson.
The scalar multiplet contains two scalars,
one Majorana spinor and two auxiliary bosons. Both sets of fields
satisfy off-shell and on-shell equilibrium between the number of
bosonic and fermionic degrees of freedom.
A field multiplet is conveniently denoted by
$(n_b,n_f,n_{aux})$, where $n_b$, $n_f$ and $ n_{aux}$  are respectively
the numbers of its propagating bosons,
propagating fermions and auxiliary boson fields,
all defined modulo gauge invariance.
The off-shell equilibrium means $n_f= n_b+n_{aux}$.
The $N=1$ vector and scalar multiplets are thus $(3,4,1)$ and $(2,4,2)$.
On a Euclidean $4$-manifold with $SU(2)$ holonomy the $N=1$ superYang-Mills
theory
can be expressed in twisted form, both for
the vector and the scalar multiplets $(3,4,1)$ and $(2,4,2)$.
This has been studied from
various points of view in \cite{{Johansen},{Witten1},{popov},{ivpo},{Hofman},{park},{park2},{bata},{recons}}.
To build the twisted formulation one
describes the spinors as holomorphic and antiholomorphic forms,
in such a way that both multiplets are decomposed as follows
\bea\label{341242}
(3,4,1) &:& A_m,  A_\bm;  \psi_m, \chi_{\bm\bn}, \chi ; h. \nonumber\\
(2,4,2)&:& \Phi, {\bar \Phi}; \psi_\bm, \chi_{mn}, { \bar \chi};
B_{\bm\bn},  T_{mn}.
\eea
Here we use the complex space coordinates $z^m, z^\bm$, where
$ m=1,2$ are $SU(2)$ indices
and $ \bm =\bar 1,\bar 2$ are the  complex conjugate ones. 
One has both a metric tensor $g_{ ij}$ and 
a complex structure  $J^ i_j$, with $J^2=-{\bf 1}$,
$J_{m \bn}= -J_{ \bn m}$, $J_{mn} = 0 =J_{\bm\bn}=0$. One can express the scalar product
in term of $J$ according to the formula $X\cdot Y= g_{ ij} X^i Y^j= iJ_{m\bn} ( X^m Y^\bn -X^\bn Y^m)$.
The raising and lowering of tensorial indices can be expressed in terms of $J$, through the formulas
$X^m = -i J^{m\bn} X_\bn$ and $Y^\bm = i J^{\bm n} Y_n$.\par
The link between the Euclidean spinors
$\lambda^\alpha,\lambda_{\dot\alpha}$ and their
holomorphic-antiholomorphic decompositions as
in Eq.~(\ref{341242}) is given by the following formula \footnote{We
define the Euclidean
$\sigma$ matrices as $\sigma_\mu=(i\tau^c,{\bf 1}_2)$,
where $\tau^c$, for $c=1,2,3$,
are the Pauli matrices.}
  \bea\label{twistn1}
\psi_m &=& \lambda^\alpha\sigma_{\mu \ \alpha\dot 1} e^\mu_m , \nn\\
\chi_{\bm\bn} &=&
\bar\lambda_{\dot\alpha}\ \bar\sigma_{\mu\nu \ \dot 2}^{\ \dot\alpha}
 \ e^\mu_\bm e^\nu_\bn  ,
\nn\\
\chi&=& \delta^{\ \dot\alpha}_{\dot 2}\bar\lambda_{\dot\alpha}.
\eea
In both sides of this equality we have a total counting of $4$ real fields.

We may consider the twist formula (\ref{twistn1}) as a mere change of variables,
such that the  Dirac
Lagrangian satisfies
\bea \label{dirac}
\bar\lambda \gamma^\mu D_\mu \lambda   &=&{\rm Tr} \Big(
-\chi^{mn}
D_{[m} \psi_{n]}   +  \chi   D^{m} \psi{_m}\Big).
\eea
The Yang--Mills Lagrangian ${\rm Tr} (F_{\mu\nu} F^{\mu\nu})$, modulo a boundary term,
satisfies
 \bea \label{ym}
 & \frac{1}{2}{\rm Tr} \Big( F_{\mu\nu} F^{\mu\nu}\Big)\sim
 {\rm Tr} \big( \frac{1}{2} F^{mn}F_{mn} + \frac{1}{4} {(F_m^m)^2}\big) \sim
   {\rm Tr} \big( \frac{1}{2}F_{mn}F^{mn} -h^2 +  h F_m^m\big),&
\eea
where $h$ is an auxiliary scalar field which can be
eliminated by a
Gaussian integration. Eqs.~(\ref{twistn1}) and (\ref{dirac}) are only
invariant under
$U(2)=SU(2)\times U(1)\subset   SO(4)$.
\par
Both  multiplets in (\ref{341242}) are the so-called
twisted expressions of the vector and scalar multiplets of the
$N=1$, $D=4$ superPoincar\'e symmetry. The conserved
{  ghost or shadow number} and the {  mass dimension} of the fields of  the
twisted multiplets are detailed in the following table
\bea
&
\begin{array}{|c|c|c||c|c|c|}\hline
fields&gh.n.&m.d.&fields:&gh.n.&m.d.\\ \hline
A_m&0&0& \Phi &2&0\\
A_{\overline m}&0&0&{\overline \Phi} &-2&0 \\ \hline
\psi_m &1&\frac{1}{2}&\psi_{\overline m} &1&\frac{1}{2} \\
\chi_{{\overline m}{\overline n}} &-1&\frac{1}{2}&\chi_{{m}{n}}
&-1&\frac{1}{2} \\
\chi &-1&\frac{1}{2}&{\overline \chi} &-1& \frac{1}{2}\\ \hline
h &0&1&T_{mn} &0&1 \\
&&& B_{mn} &0&1 \\ \hline
\end{array}
&
\eea
One defines the  four twisted supersymmetry operators $\s,s_\bp$ and $s_{qr}$
by the following algebra,  with the possibility of
  gauge transformations of
the r.h.s.,
\bea
 \{ \s, s_{\overline p} \} &=& \partial_{\overline p}  +
 \gauge(A_{\overline p}),\nonumber\\
 \{ s_{\overline p}, s_{qr} \} &=& a
 J  _{\bp[q }  (  \partial_{r]}   +\gauge(A_{r]}))
\eea
(the remaining anticommutators are all vanishing, which means that,
in particular, each one of the operators is nilpotent).
The above algebra (with $a$ an arbitrarily given dimensionless constant)
is the most general one which is compatible with the various charge assignments.
Its generators anticommute with
$\partial=dz^a\partial_{z^a}$ and
${\bar \partial}=d{\overline z}^a\partial_{{\overline z}^a}$.
Their charge assignments are
\bea
&\s\equiv (gh.n=1, m.d.=\frac{1}{2}),
\quad s_{\overline p}\equiv (gh.n=-1, m.d.=\frac{1}{2}), \quad s_{qr}\equiv (gh.n=1, m.d.=\frac{1}{2}).&\nonumber\\
\eea
Without loss of generality we can consider two cases,
either $a=0$ or $a=1$. The $a=0$ case is degenerate. It implies
not using the K\"ahler metric. Power counting allows one to compute the field
transformations that are compatible with this algebra. They are given by
{\footnotesize \bea
&
\begin{array}{|c||c|c|c|}\hline
&\s&s_{\overline p}&s_{qr}\\ \hline\hline
A_m&\psi_m&a    J_{\bp m }\chi &0\\
A_\bm&0&\chi_{\bp \bm }&a J_{\bm [r }   \psi_{q] }\\ \hline
\psi _m&0&F_{\bp m}  - a   J_{\bp m}  h &0\\
\chi&h&0&D_{[q} \psi_{r]}\\
\chi_{\bm\bn}&F_{\bm\bn}&0&
\begin{array}{c}
-a  (J_{\bm [r }F_{q]\bn }
-  J_{\bn [r }F_{q]\bm })\\
+\frac{a^2}{2} (J_{\bm [r }J_{q]\bn }
-  J_{\bn [r }J_{q]\bm })h
\end{array}
\\ \hline
h&0&D_\bp \chi&D_{[q} \psi_{r]}\\ \hline\hline
\Phi&0& -\psi  _ \bp&0\\
\bar \Phi &\bar  \chi& 0&-a   \chi_{qr }\\ \hline
\psi_\bm &-D_\bm \Phi&B_{\bp \bm }&a      J_{\bm [ q } D_{ r]}\Phi \\
\chi_{mn} &T_{mn}&2   J_{\bp [ m }  D_{ n]}\bar \Phi&0 \\
\bar \chi &0&D_\bp \bar\Phi&a \chi_{qr } \\ \hline
T_{mn} &0&-2  J_{\bp [m }    D  _{n]}\bar \chi  +D_\bp \chi_{mn}
-2  J_{\bp [m }  \bar\Phi\cdot \psi_{n ]} &0 \\
B_{\bm \bn}&2D_{[\bm }  \psi _{ \bn]}+\chi_{\bm\bn}\cdot\Phi   &0&a
(
  J_{\bm [ q } D_{r]} \psi_{\bn}-
 J_{\bn [ q } D_{r]} \psi_{\bm}
 )  \\ \hline
\end{array}
&\nonumber\\
\eea
}
The non-trivial case is $a=1$. In this case one can associate to $\s,
 s_{\overline p}, s_{qr}$ the four operators $Q^\alpha, Q_{\dot \alpha}$, with a relation as
Eq.~(\ref {twistn1}). One then finds that the following  supersymmetry algebra for the $Q$'s,

\bea
\{  Q^\alpha, Q_{\dot \beta } \} = {\sigma^\alpha _{\dot \beta }}^\mu \partial_\mu
\eea
modulo gauge transformations, establishes a link between the $N=1$, $D=4$
superPoincar\'e algebra and its twisted version.   \par
The twisted transformation laws of $(3,4,1)$ are independent
from those of $(2,4,2)$, while the converse is true only in the abelian limit,
the coupling being due to the transformation laws of
the auxiliary fields $ B_{\bm\bn}$ and  $T_{mn}$.
One should also note that for $a=0$ the tensor symmetry
$s_{qr}$ is completely degenerated for $(2,4,2)$, but not
for the $(3,4,1)$ twisted multiplet.\par
The most general $\s$ and $s_\bp$ invariant actions
of second order in the derivatives of the bosonic fields correspond to both Lagrangians
\bea\label{twisteda}
{\cal L}_{341}&=& \s \ {\rm Tr} \Big(\frac{1}{2}\chi^{mn}F_{mn} +\chi (-h +F_m^m) \Big),\nonumber\\
{\cal L}_{242}&=&\s \ {\rm Tr}\ \Big(\frac{1}{2}\chi^{\bm \bn}B_{\bm\bn}+  \bar \Phi
(D^\bm \psi_\bm   +  \Phi\cdot \chi)\Big).
\eea
They are nothing else than the $N=1$ $D=4$
Lagrangians,  as can be verified by  computing the  $s$-exact terms and using both  equations~(\ref{dirac}) and
~(\ref{ym}) \cite{recons}.\par
One can then check that $s_{qr}$ is also a symmetry of both actions
${\cal L}_{341}$ and $
{\cal L}_{242}$.
The  $s_{qr}$ symmetry is thus a redundant symmetry, which is already determined from
the invariance under the three generators
$\s$ and $s_\bp$. Such redundant symmetries are also present in higher dimension. They often only close when using the equations of motion.
On the other hand, in all studied cases the non-redundant symmetries, which uniquely determine the action, close off-shell.  For instance,
the anticommutation relations of the redundant $ s_{qr}$ symmetry in the $SU(4) \subset SO(8)$
decomposition of
$N=2$, $D=8$ with the $(9,16,7)$ multiplet
close only on-shell [work in  preparation].

The introduction of the fourth generator $s_{qr}$ is however
necessary to untwist the $4$
generators $\s,s_\bp, s_ {qr}$
into the $4$ superPoincar\'e generators $Q^\alpha, Q_{\dot\alpha}$.
We will actually investigate how the twisted and untwisted formulations
are related in one dimension and we will use 
the existence of the fourth supersymmetry generator.

\section {The gauge $A_{\bar 2}=0$.}

One may wish to directly understand the balance between bosons and
fermions
without referring to gauge invariance by reducing the number
of degrees of freedom of the gauge field. This implies a breaking of the
gauge symmetry,
which turns out to imply a dimensional reduction.\par
We can for instance set $A_{\bar 2}=0$. Further constraints are required
to maintain the
supersymmetry algebra. Indeed, the equations
\bea
s_{\bar 1} A_{\bar 2} &=& \chi _{\bar 1 \bar 2},\nonumber\\
s_{qr} A_{\bar 2} &=&  a J_{\bar 2 r }   \psi_{q }
\eea
cannot be preserved when imposing $A_{\bar 2}=0$.
It follows that, in order to preserve the consistency
of the subalgebra  of the  $3$ generator $\s$ and $s_\bp$
acting on the vector multiplet $(3,4,1)$,
one must set to zero all fields derivatives with
respect to the variable $z^{\bar 1}$. One is reduced to a theory in
$3$ dimensions with 4 bosons, $ A_m, A_ {\bar 2}$ and $ h$,
the non-nilpotent part of the algebra
being
\bea
\{ s, s_{\overline 1} \} &=& \gauge_{{\rm global}}(A_{\overline 1}),\nonumber\\
\{ s, s_{\overline 2} \} &=& \partial_{\overline 2} .
\eea
In this theory the $SU(2)$ covariance has disappeared. \par
In order to maintain the full $4$-generator algebra
(including $s_{qr}$) the constraint
is more drastic. As indicated by the $s_{qr}$-transformation
law of
$\chi _{\bar 1 \bar 2}$, one
  must set to zero all field derivatives with
respect to the variable $z^{m}$, for $m=1,2$, to preserve
the equation
$s_{qr} A_{\bar 2} =  a J_{\bar 2 r }  \psi_{q }$.
One then recovers a one-dimensional theory
with new features that will be detailed
in the next Section.
The algebra now reads
 \bea
 \{ s, s_{\overline 1} \} &=&
 \gauge _{{\rm global}}(A_{\overline 1}),\nonumber\\
  \{ s, s_{\overline 2} \} &=& \partial_{\overline 2},   \nonumber\\
  \{ s_{\overline p}, s_{qr} \} &=&
  a     J  _{\bp[q }  (   \gauge_{{\rm global}}(A_{r]})).
 \eea
Modulo global gauge transformations, its only non trivial anticommutator is
$ \{ s, s_{\overline 2} \}$, which
  expresses the one-dimensional supersymmetry.

\section{Twisted supersymmetry in one-dimension.}

The gauge-fixing $A_{\overline 2}=0$ implies that the
twisted superalgebra
with the  four generators
$\s, s_{\overline p}$ and $ s_{qr}$ is defined in $D=1$ dimension and
closes modulo the (remnants of the) gauge transformations.
We can therefore analyze the differences between the $D=1$ dimensional
reductions of twisted and untwisted supersymmetries, for both the
$(3,4,1)$ and $(2,4,2)$ multiplets.
\par
Modulo gauge transformations, the transformations of the
dimensionally reduced twisted vector multiplet $(3,4,1)$ are given by
\bea
&
\begin{array}{|c|c|c|c|c|}\hline
&s&s_{\overline 1}&s_{\overline 2}& s_{12}\\\hline
A_1&\psi_1&0&-a\chi&0\\
A_2&\psi_2&a\chi&0&0\\
A_{\overline 1}&0&0&-\chi_{{\overline 1}{\overline 2}}&a\psi_1\\
\psi_1&0&0&{\dot A}_1+ah&0\\
\psi_2&0&-ah&{\dot A}_2&0\\
\chi_{{\overline 1}{\overline 2}}&-{\dot A}_{\overline 1}&0&0&a({\dot A}_1+ah)\\
\chi&h&0&0&0\\
h&0&0&{\dot \chi}&0\\ \hline
\end{array}
&
\eea
The mixed transformations can be eliminated by field
redefinitions that  break the shadow number. Such a
breaking is admissible because a tensorial index becomes an internal index in  $D=1$.
We can express the field redefinitions as
\bea
\begin{array}{lll}
z_1 = A_1, &\quad &\xi_1 = \psi_1-a\chi,\\
z_2= A_{\overline 1} -a A_2,&\quad &\xi_2 = \psi_1+a\chi,\\
z_3 = A_{\overline 1} +a A_2,&\quad &\xi_3 = \chi_{{\overline 1}{\overline 2}},\\
g= {\dot A}_1+2ah, &\quad &\xi_4 = \chi_{{\overline 1}{\overline 2}}-a \psi_2.
\end{array}
\eea
By introducing the basis of four operators
\bea
s_\pm = \s\pm s_{\overline 2}, &\quad&
N_\pm = \frac{1}{a}(s_{12}\pm s_{\overline 1}),
\eea
one obtains a realization of the ${\cal N}=[1,1,2]$
generalized supersymmetry. The
four operators $s_\pm$ and $ N_\pm$ are indeed mutually anticommuting and satisfy
\bea
{s_\pm}^2=\pm \partial_t  &,& {N_\pm}^2 = 0.
\eea
No mixed term occurs  for  the twisted $(2,4,2)$ multiplet. If  one defines
\bea &
\begin{array}{lll}
w_1 = {\overline \Phi}, &\quad & \mu_1 = {\overline \chi},\\
w_2 = \Phi, &\quad & \mu_2 = \psi_{\overline 2},\\
d= T_{12}, &\quad &\mu_3 = {\overline \chi}_{12},\\
f= B_{{\overline 1}{\overline 2}}, &\quad &\mu_4 = -\psi_{\overline 1},
\end{array}&
\eea
one finds that,  for the $a=1$ case, the $(3,4,1)$ and $(2,4,2)$
twisted  multiplets transform as follows
\bea
\begin{array}{|c|c|c|c|c|}\hline
&s_+&s_-&N_+& N_-\\\hline
z_1&\xi_1&\xi_2&0&0\\
z_2&-\xi_3&\xi_4&\xi_1&\xi_2\\
z_3&-\xi_4&\xi_3&-\xi_2&\xi_1\\
\xi_1&{\dot z}_1&-g&0&0\\
\xi_2&g&-{\dot z}_1&0&0\\
\xi_3&-{\dot z}_2&-{\dot z}_3&{\dot z}_1&g\\
\xi_4&-{\dot z}_3&-{\dot z}_2&g&{\dot z}_1\\
g&{\dot \xi}_2&{\dot \xi}_1&0&0\\ \hline
\end{array}
&\quad,\quad& \begin{array}{|c|c|c|c|c|}\hline
&s_+&s_-&N_+& N_-\\\hline
w_1&\mu_1&\mu_1&-\mu_3&-\mu_3\\
w_2&\mu_2&-\mu_2&-\mu_4&\mu_4\\
\mu_1&{\dot w}_1&-{\dot w}_1&d&d\\
\mu_2&{\dot w}_2&{\dot w}_2&f&-f\\
\mu_3&d&d&0&0\\
\mu_4&f&-f&0&0\\
d&{\dot \mu}_3&-{\dot \mu}_3&0&0\\
f&{\dot \mu}_4&{\dot \mu}_4&0&0\\ \hline
\end{array}
\eea
Let us stress that the disappearance of the mixed terms allows one  to present the twisted
transformations in a graphical form, in analogy with the untwisted
$D=1$ supersymmetric case~\cite{fg}. \par
It is natural to examine whether the above transformations can be recovered in terms of linear combinations of
the   ${\cal N}=[3,3,0]$
pseudosupersymmetry   defined in~\cite{pt}, by computing its action on the $(3,4,1)$ and $(2,4,2)$ multiplets, respectively.
The ${\cal N}=[3,3,0]$ pseudosupersymmetry is made of  the 6 operators $Q_i$, ${\overline Q}_i$, $i=1,2,3$,
such that ${Q_i}^2=\partial_t$, ${{\overline Q}_i}^2=-\partial_t$,  and they all
   mutually anticommute.
The  ${\cal N}=[3,3,0]$
pseudosupersymmetry transformations can be
written as follows\bea
\begin{array}{|c|c|c|c|c|c|c||c|}\hline
&Q_1&{\overline Q}_1&Q_2& {\overline Q}_2&Q_3&{\overline Q}_3&Q_4\\\hline
z_1&\xi_1&\xi_1&-\xi_4&-\xi_4&\xi_2&\xi_2&\xi_3\\
z_2&-\xi_3&\-\xi_3&-\xi_2&\xi_2&-\xi_4&\xi_4&\xi_1\\
z_3&-\xi_4&\xi_4&-\xi_1&\xi_1&\xi_3&\xi_3&-\xi_2\\ \hline
\xi_1&{\dot z}_1&-{\dot z}_1&-{\dot z}_3&-{\dot z}_3&-g&-g&{\dot z}_2\\
\xi_2&g&g&-{\dot z}_2&-{\dot z}_2&{\dot z}_1&-{\dot z}_1&-{\dot z}_3\\
\xi_3&-{\dot z}_2&{\dot z}_2&-g&g&{\dot z}_3&-{\dot z}_3&{\dot z}_1\\
\xi_4&-{\dot z}_3&-{\dot z}_3&-{\dot z}_1&{\dot z}_1&
-{\dot z}_2&-{\dot z}_2&-g\\ \hline
g&{\dot \xi}_2&-{\dot\xi}_2&-{\dot\xi}_3&-{\dot\xi}_3&
-{\dot\xi}_1&{\dot\xi}_1&-{\dot \xi}_4\\ \hline
\end{array} &\quad,\quad&
\begin{array}{|c|c|c|c|c|c|c||c|}\hline
&Q_1&{\overline Q}_1&Q_2& {\overline Q}_2&
Q_3&{\overline Q}_3&Q_4\\\hline
w_1&-\mu_3&-\mu_3&\mu_2&\mu_2&\mu_1&\mu_1&\mu_4\\
w_2&\mu_4&\mu_4&-\mu_1&\mu_1&\mu_2&-\mu_2&\mu_3\\ \hline
\mu_1&d&d&-{\dot w}_2&-{\dot w}_2&{\dot w}_1&-{\dot w}_1&-f\\
\mu_2&-f&-f&{\dot w}_1&-{\dot w}_1&{\dot w}_2&{\dot w}_2&-d\\
\mu_3&-{\dot w}_1&{\dot w}_1&-f&-f&d&d&{\dot w}_2\\
\mu_4&{\dot w}_2&-{\dot w}_2&d&-d&f&-f&{\dot w}_1\\ \hline
d&{\dot\mu}_1&-{\dot\mu}_1&{\dot\mu}_4&{\dot\mu}_4&{\dot\mu}_3&
-{\dot\mu}_3&-{\dot \mu}_2\\
f&-{\dot\mu}_2&{\dot\mu}_2&-{\dot\mu}_3&{\dot\mu}_3&{\dot\mu}_4&
{\dot\mu}_4&-{\dot \mu}_1\\ \hline
\end{array}
\eea
Here a  seventh column has been added   to express the action of the
fourth generator
$Q_4$, of  the larger ${\cal N}=[4,0,0]$ supersymmetry. The action of $Q_4$ is in fact
fixed, up to an overall sign, once  the transformations of the fields under     $Q_1, Q_2$ and  $Q_3$ have been determined
\footnote{The dimensional reduction of the untwisted vector and matter
multiplets must produce the $(3,4,1)$ and $(2,4,2)$ linear
representations of the Euclidean ${\cal N}=[4,0,0]$ superalgebra~(\ref{gensusy}) \cite{{wb},{top}}.}.
\par
For the $(3,4,1)$ twisted multiplet one can identify
\bea
&s_+\equiv  Q_1,\quad
s_- \equiv {\overline Q}_3,\quad
N_- \equiv  \frac{1}{2}({\overline Q}_2-Q_2).&
\eea

\par
Similarly, a $3$-generator subalgebra of the $(2,4,2)$ twisted
multiplet can be embedded into the
 ${\cal N}=[3,3,0]$ pseudosupersymmetry through
\bea\label{reconstruction}
&s_+\equiv Q_3,\quad
s_-\equiv {\overline Q}_3,\quad
N_-\equiv \frac{1}{2}(Q_1+{\overline Q}_1).&
\eea
 The operator $N_+$, on the other hand, cannot be obtained as a
linear combination of the $Q_i$'s and ${\overline Q}_i$'s, for  both   the    $(2,4,2)$ and  $(3,4,1)$  twisted mutiplets.  \par
It is worth noticing that, for the $(2,4,2)$ multiplet,  the
subalgebra generated by $s_\pm, N_-$ can be recovered as a
subalgebra of the ${\cal N}=[2,2,0]$ pseudosupersymmetry,
since only $Q_1,Q_3, {\overline Q}_1$ and $ {\overline Q}_3$ enter
Eq.~(\ref{reconstruction}).\par
An important remark is that the twisted matter multiplet
$(2,4,2)$ {  cannot} be recovered from a dressing \cite{{pt},{krt}} of
the twisted gauge multiplet $(3,4,1)$ (by identifying the
extra auxiliary field with either ${\dot z}_1$, ${\dot z}_2$
or ${\dot z}_3$). \par
To summarize,  the $3$-generator subalgebra made of
$\s$ and $ s_{\overline p}$ induces, after  suitable redefinitions
of the generators, an ${\cal N}=[1,1,1]$ generalized supersymmetry.
The   $4$-generator algebra obtained with the addition of
  $s_{qr}$   induces in $D=1$
an ${\cal N}=[1,1,2]$ generalized supersymmetry.
One could wonder whether the representations of the ${\cal N}=[1,1,2]$
generalized supersymmetry on the twisted multiplets
could be recovered from the known ${\cal N}=[3,3,0]$ pseudosupersymmetry
representations. This is not the case. At most, one can embed a
$3$-generator subalgebra ${\cal N}=[1,1,1]\subset {\cal N}=[1,1,2]$
into the ${\cal N}=[3,3,0]$ pseudosupersymmetry.
This ${\cal N}=[1,1,1]$ subalgebra, such that
${\cal N}=[1,1,2]\supset {\cal N}=[1,1,1]\subset {\cal N}=[3,3,0]$,
is  generated by $\s, s_{\overline 2}$ and a linear combination of $s_{12}$ with $s_{\overline 1}$.
It should be stressed that the inequivalent ${\cal N}=[1,1,1]\subset {\cal N}=[1,1,2]$ subalgebra generated by
$\s, s_{\overline 2}$ and $ s_{\overline 1}$ is not
contained in $N=[3,3,0]$. An explicit computation proves that, for both these $N=[1,1,1]\subset N=[1,1,2]$ embeddings, each one  of  these   $3$-generator invariances fixes the same Lagrangian, and thus determines the
full $N=[1,1,2]$ $4$-generator invariance.\par
As a final remark, let us    mention that the twisted and untwisted
supersymmetry
in $D=1$ can be regarded as acting on the same set of component
fields. However, they   only
admit    one common generator.

\section{Conclusions and outlook.}

The algebra of the Poincar\'e supertranslations is given by $4$ generators. In
a twisted form, however, a $3$-generator subalgebra
is sufficient
to determine the invariant actions. The consistency of the gauge-fixed
$3$-generator
subalgebra produces a $D=3$ theory. On the other hand, the consistency
of the full
gauge-fixed $4$-generator algebra induces a one-dimensional theory. We have seen that
the gauge-fixed
one-dimensional reduction of the twisted supersymmetry differs from the
dimensional reduction
of the untwisted supersymmetry. The latter is a supersymmetry with
${\cal N}=[4,0,0]$ supercharges
(their squares are positive and coincide with the Hamiltonian),
while the former is a ${\cal N}=[1,1,2]$ generalized supersymmetry,
where one operator has a positive square, one operator has a negative square and
the two remaining ones are nilpotent.
This has been checked on the vector and the scalar multiplets of the
$N=1$, $D=4$ theory.
In this example the untwisted and
twisted
supersymmetry only share one
common generator and are not equivalent,
in the usual sense of superalgebra representations, although a
complexification, followed by a linear mapping and
a reality condition, suggests their very close link, that can be called the twist.
We have shown that the twisted ${\cal N}=[1,1,2]$ supersymmetry acting on the vector
and the
matter multiplets {cannot} be obtained as a result of an embedding into an
${\cal N}=[3,3,0]$ pseudosupersymmetry. In contrast, the $3$-generator algebras
${\cal N}=[3,0,0]$ and ${\cal N}=[1,1,1]$, which
are two distinct subalgebras of the ${\cal N}=[3,3,0]$ pseudo-supersymmetry,
can be obtained.
They admit the same invariant Lagrangian, modulo field redefinitions.

Extending the present investigation to the dimensional
reduction of the $N=2$ $D=4$
twisted SuperYang-Mills theory is rather straightforward.
It produces a
one-dimensional twisted supersymmetry
realized on the $(5,8,3)$ set of fields.
Much more interesting is the application to the twisted version of
the $N=4$
superYang--Mills theory with its $(9,16,7)$ multiplet.
The latter theory is conformally-invariant
and, in its planar limit, integrable hierarchies
are recovered. The most relevant part consists of a twisted
$6$-generator subalgebra which closes off-shell and uniquely
determines the theory. The dimensional reduction to low dimensions
($D=1,2$) allows one  to use such powerful tools like
the Lax pairs to analyze the integrable properties of the theory.
We leave this investigation for future works.\\{}~
\par
{\large{\bf Acknowledgments}}{} ~\\{}~\par

F.T. is grateful to the LPTHE for the hospitality.
The work was partially supported by Edital Universal CNPq,
Proc. 472903/2008-0.

\end{document}